\documentclass[twocolumn,american,english]{revtex4}
\usepackage[T1]{fontenc}
\usepackage[latin9]{inputenc}
\usepackage{graphicx}
\usepackage{amssymb}

\makeatletter

\usepackage{babel}
\makeatother

\begin{document}

\title{From protein binding to pharmacokinetics: a novel approach to active
drug absorption prediction.}

\author{P.O. Fedichev, T.V. Kolesnikova, and A.A. Vinnik}

\affiliation{Quantum Pharmaceuticals, Ul. Kosmonavta Volkova 6-606, 125171, Moscow,
Russia}

\selectlanguage{american}%
\begin{abstract}
Due to inherent complexity active transport presents a landmark hurdle
for oral absorption properties prediction. We present a novel approach
carrier-mediated drug absorption parameters calculation based on entirely
different paradigm than QSPR.\foreignlanguage{english}{ We capitalize
on recently emerged ideas that }molecule activities against a large
protein set can be used for prediction of biological effects\foreignlanguage{english}{
and performed a large scale numerical docking of drug-like compounds
to a large diversified set of proteins. As a result we identified
for the first time a protein, binding to which correlates well with
the intestinal permeability of many actively absorbed compounds. Although
the protein is not a transporter, we speculate that it has the binding
site force field similar to that of an important intestinal transporter.
The observation helped us to improve the passive absorption model
by adding non-liner flux associated with the transporting protein
to obtain a quantitative model of active transport. This study demonstrates
that binding data to a sufficiently representative set of proteins
can serve as a basis for active absorption prediction for a given
compound.}
\end{abstract}
\maketitle
\selectlanguage{english}%

\section{Introduction}

The oral route of drug administration is very convenient for patients,
however it is often inefficient due to low solubility, intestinal
permeability, or high first-pass effect. Therefore prediction of oral
absorption properties is of great interest for pharmaceutical industry.
Orally administered drugs are mainly absorbed in the small intestine.
Here, depending on drug composition and size, absorption can happen
through a variety of processes \citep{hunter1997isd}. Drug pass through
the epithelial cells and the lamina propria from the lumen into the
blood stream in the capillaries. On its way it might be metabolized,
transported away from the tract where absorption is possible or accumulate
in organs other than those of treatment. Besides a fundamental interest
in understanding the basic mechanisms by which a drug is assimilated
by the human body, the kinetics of drug absorption is also a topic
of much practical interest. A detailed knowledge of this process,
resulting in the prediction of the drug absorption profile, can be
of much help in the drug development stage \citep{eddershaw2000app}. 

There are a number of kinetic absorption models were developed that
require experimentally determined intestinal permeability of a compound
as an input \citep{yu1999caa}. Although of great value such {}``hybrid''
partially experimental, partially computational models miss the main
advantages of purely theoretical approaches: no need in chemical synthesis
of a compound and experimental facilities, low cost and high speed.
Among computational approaches that predict intestinal permeability
solely from a molecule structure and its physical-chemical properties
instead of using any biological experiments data, there are two major
directions: ab initio and quantitative structure-property relationship
(QSPR) models. The last ones are overwhelmingly used nowadays and
exploit a wide spectrum of statistical methods for absorption data
analysis (see e.g. \citep{hou2006rac,bergstrom2005spd,subramanian2006cam}
for a review). Instead of relying on basic laws of nature the models
are trained at observed statistical regularities. Such an approach
preconditions the limitations of the models. In contrast to QSPR there
are a handful of studies developing models of the intestinal permeability
from the first principles \citep{adson1995pdw,camenisch1998smp,obata2005pod}.
The models describe successfully basic properties of passive absorption:
dependence on distribution coefficient, diffusional limitation at
high $LogD$, and paracellular absorption. However, the major hindrance
on this way the complexity of intestinal absorption. Apart from passive
phenomena (diffusion through cell membrane and paracellular junctions),
there is also active transport of the molecules in and out of the
cells. To the best of our knowledge current ab initio models are limited
to description of drug passive absorption. Most of QSPR models also
deal with passive transport \citep{bergstrom2005spd}, though only
a few approaches go as far as developing QSPR models describing both
passive and carrier-mediated absorption mechanisms \citep{wessel1998phi}.
However, carrier-mediated transport plays an important role in drug
absorption \citep{dobson2008cmc} and hence demands the development
of a good active absorption model.

\selectlanguage{american}%
The major objective of this investigation was to develop a novel approach
to prediction of carrier-mediated drug absorption based on entirely
different paradigm than QSPR, thus avoiding its difficulties and capable
of better predictions. Recently it was observed that experimental
values of molecular activities against a large protein set can be
used for prediction of a broad spectrum of biological effects . In
this study we took advantage of this concept and developed a novel
quantitative method for identification of actively transported drugs.
To do that we performed a docking study of a few hundreds of small
molecules (mostly drugs) against a diversified set of $400$ proteins
representing human proteom. \foreignlanguage{english}{Using available
absorption data for each of the molecules we identified a protein,
affinity for which correlates well with the permeability of many actively
absorbed compounds from our data set. The observation helped us to
improve the passive absorption model by adding non-liner fluxes associated
with the transporting protein to obtain also a quantitative model
of active transport.}

\selectlanguage{english}%
The manuscript is organized as follows. After the standard Materials
and Methods section outlining our approaches to the data preparation,
the docking study setup, and the data processing routines, we present
a two-compartment model of drug absorption extended to include active
transport via non-linear fluxes terms associated with transporting
proteins. As soon as the model is built and the parameters of passive
absorption are fitted to experimental data, we identify the active
transport parameters to train the classifier. After the classification
is set up we compare our predictions with available experimental informaton
and thus validate the complete model for drug absorption prediction.

\section{Materials and Methods}

\subsection{Experimental absorption and permeability data }

Much experimental activity aimed to analyze the kinetic aspects of
the process of drug absorption has been pursued recently. For better
control, a variety of in-vitro methods on drug absorption have been
developed \citep{balimane2000cmu}. One possibility is to seed (epithelial)
cell cultures in a mono-layer, forming the contact surface of two
little pots. Concentrations of an applied drug can be measured over
time in both chambers. Two well known cell culture models are Caco-2
cells \citep{artursson1997ida,artursson2001cme} and MDCK cells \citep{irvine1999mmd}.

\begin{figure}
\includegraphics[width=0.9\columnwidth]{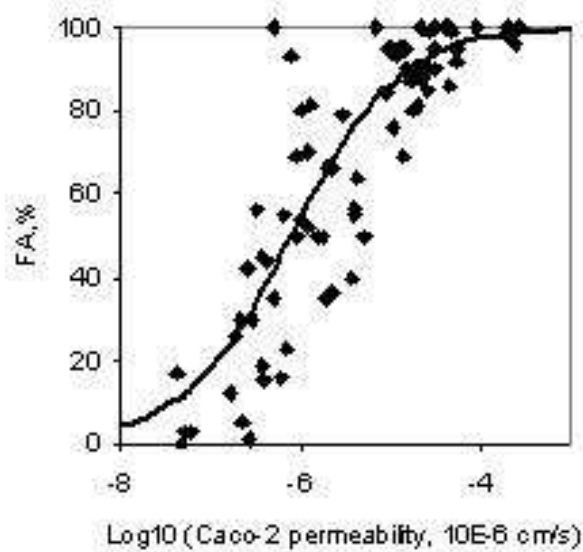}\caption{The relationship between FA and Caco-2 permeability. The points corresponds
to experimental values for compounds with both values of $FA$ and
$P$ known. The solid line is the approximation provided by Eq. \ref{eq:FA(P)}.
RMSD is 14\%.\label{fig:FA(P)} }

\end{figure}

To enrich experimental data sets we used two types of observed data
to build up the model: fraction of drugs absorbed after oral administration
in humans ($FA$) and permeability across a human colon adenocarcinoma
cell (Caco-2) monolayer ($P$). The latter is a routinely used cell
model in pharmaceutical industry and academia to estimate drug absorption
in the intestine \citep{artursson2001cme,ingels2003bpa,vandewaterbeemd2001pbd,balimane2006cip}).
Previous findings showed strong relationship between drug Caco-2 permeability
and the fraction absorbed in humans (e.g. \citep{artursson1991cbo,yee1997vpa,tavelin1999cie,stenberg2001eac,balimane2006cip,matsson2005erd}),
suggesting that one value can be used to estimate the other. We collected
from literature compilations $91$ observed $FA$ values and $103$
Caco-2 permeability values for $117$ compounds that to the best of
our knowledge are not subject to efflux from enterocytes \citep{parrott2002pia,zhao2002rls,klopman2002aec,agatonovickustrin2001tdm,stenberg2001eac,raevsky2000qed,ghuloum1999mhn,clark1999rcp,oprea1999tmm,wessel1998phi,palm1997pms,ertl2000fcm,martinez2004cnn,sanghvi2003ppi,kansy1998pht,linnankoski2006cpo,nordqvist2004gmp,gunturi2007sam,balon1999dlp,yazdanian1998cpa,raevsky2002nap,guangli2006pcp,castillogarit2007eap,liang2000mta,subramanian2006cam,ponce2004ntd,katsura2003iad,yee1997vpa,tavelin1999cie,matsson2005erd,camenisch1998epp,balimane2006cip,Fish2007ddi,vanbambeke2003aep,hou2007aed1,hou2007aed2}.
Fig. \ref{fig:FA(P)} shows $FA$ values plotted against permeability
for the compounds, for which both values were available. The data
were fitted with the sigmoid equation \citep{artursson2001cme}: \begin{equation}
FA=\frac{100\%}{1+(P/P_{50})^{p}}\label{eq:FA(P)}\end{equation}
where $P_{50}$ is the permeability at $50\%\, FA$, and $p$ is a
slope factor. The fitting parameters were $P_{50}=7.94\times10^{-7}$
${\rm cm\cdot s^{-1}}$ and $p=-0.73$ that is in reasonable agreement
with previously found $P_{50}\sim2\times10^{-6}$ ${\rm cm\cdot s^{-1}}$
and $p=-0.5$ \citep{matsson2005erd}. The fitting curve predicts
$FA=90\%$ for $logP_{90}=-4.8$ and $FA=10\%$ for $logP_{10}=-7.4$,
which is in a reasonable agreement with with $logP_{90}=-5.3$ and
$logP_{10}=-6.9$ from \citep{stenberg2001eac}. RMSD of the fitting
is fairly small and thus Caco-2 permeability can indeed predict human
intestinal absorption of orally taken drugs with reasonable accuracy.
Fig. \ref{fig:FA(P)} shows that there are two outliers corresponding
to glycylsarcosine and amoxicillin. Their $FA$ were much higher than
expected from Caco-2 permeability. Glycylsarcosine and amoxicillin
are carried through enterocyte membranes by PEPT transporters, which
are reported to have reduced activity in Caco-2 cells \citep{lennernas1996cba,chong1996vpt}.
This fact may account for observed discrepancy between measured Caco-2
permeability and $FA$ values. %

Eq. \ref{eq:FA(P)} can be used to estimate the missing values of
$FA$ and $P$ for all the compounds from our compilation. However,
Eq. \ref{eq:FA(P)} requires that if $FA\rightarrow100\%$, then $P\rightarrow\infty$.
Therefore, if the observed value of $FA$ exceeded $97\%$, we assigned
$P$ value of $4\times10^{-4}$ ${\rm cm/s}$ corresponding to $97\%$
$FA$.

The distribution coefficients, $LogD$ ($pH=7.4$) used throughout
the research, were either collected from literature \citep{yamazaki2004cpp,matsson2007gdi,camenisch1998epp}
or calculated using Quantum software version 3.3.0 \citep{QUANTUM}.

\subsection{Preparation of the protein panel}

\selectlanguage{american}%
Our protein data set includes $400$ proteins form the Protein Data
Bank \citep{PDB:Westbrook2003}. It covers about almost all available
cytoplasmic proteins with known $X$-ray structure and also includes
some important transmembranal proteins such as ion channels and GPCRs.
We use homology models for GPCRs since no experimentally determined
structure is available \citep{schwede2003sma}.

Only the proteins that are co-crystallized with biologically active
ligand were taken to the data set. Ligands may be either natural ligands
(such as hormone for a hormone receptor or substrate for enzyme),
or drugs, inhibitors etc. If there exist multiple files in PDB repository
for the same protein, we consider the file with the most complete
structure and/or the lowest resolution. 

Although the choice of the proteins for the calculations is a very
important step and the overall number of proteins is hardly manageable,
we believe that the PDB archive contains a representative set of the
most practically important proteins, covering the whole interesting
variety of ligand binding domains. Below we show, that successful
predictions do not require the presence of a specific ligand binder
in the protein set employed for the calculations. Instead, it proves
to be sufficient to have a structurally similar protein in the protein
panel.

\selectlanguage{english}%

\subsection{Docking setup and the binding constant, $K_{d}$, prediction.}

\selectlanguage{american}%
Both the proteins and small molecules typization, and \emph{in-silico}
screening were carried out by the molecular processing and docking
tools taken from the QUANTUM drug discovery software suit \citep{QUANTUM}.
The software predicts the binding affinities of small molecules to
resolved protein targets using a set of first principles based molecular
simulations with an advanced continuous water model \citep{fedichev-2006,fedichev2008fep}.
The approach provides the logarithmic values of the binding constant,
$pKd$ ($-\lg Kd$). 

To compute the binding affinities of molecules in our data set we
screened each of the molecules against every protein in our panel.
To speed up the calculations the docking run were performed against
rigid protein structures with no further refinement by molecular dynamics.
Such a simplified approach turned out to be sufficient (see the discussion
below) and the results of the calculations were organized into screening
assays containing $pKd$ values for each protein-small molecules pair
(complexes) and were stored for further analysis.

\selectlanguage{english}%

\subsection{Data processing and modeling.}

Fitting of the experimental data to the models proposed below was
performed using BFGS algorithm implemented in in-house program. Selection
of proteins, affinity for which correlates with active absorption
was performed using Weka v.3.5 data mining software \citep{WEKA:witten2005dmp}.

\section{Results}

\subsection{The model}

\begin{figure}
\includegraphics[width=0.9\columnwidth]{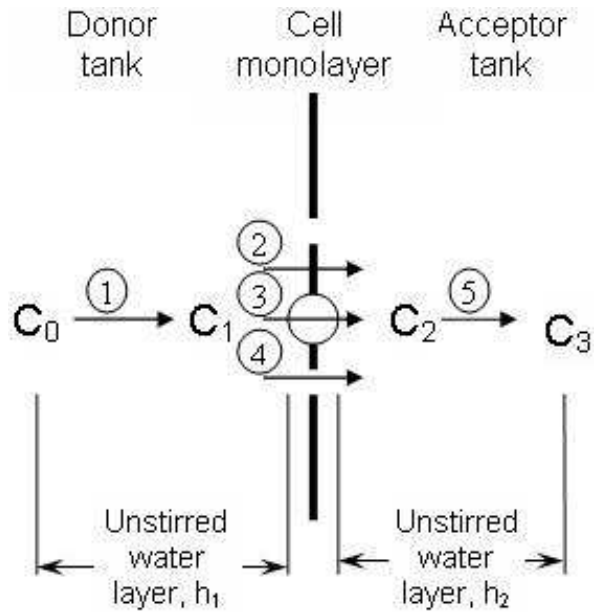}

\caption{Model of absorption used in the study. The figures represent: $1$
and $5$ \textendash{} drug diffusion from the balk solution of the
donor tank to enterocytes and from enterocytes to the balk solution
of the acceptor tank; $2$ and $3$ \textendash{} passive and active
penetration through cells; $4$ \textendash{} drug diffusion inside
enterocyte from the apical to basal membrane of enterocytes; $4$
\textendash{} paracellular absorption of the drug. The drug dissolving
stage in the intestinal lumen is omitted.\label{fig:two_comp_model}}

\end{figure}

For the sake of simplicity we considered absorption of passively and
actively transported drugs with negligible efflux and intestinal metabolism.
Besides, the model assumes that the drug is good soluble and stable
in the gastrointestinal fluids, and absorption on intestinal content
and intestinal metabolism are negligible. In this case the absorption
from intestinum to blood can be represented by a two-compartment model
(see Fig. \ref{fig:two_comp_model}) consisting of a donor (intestinal
lumen) and an acceptor (blood vessel) tanks. The intestinal wall can
be represented by a single lipid membrane since there is no phenomena
depending on drug concentration in enterocytes. The drug absorption
can be described as drug diffusion from the balk solution of the donor
tank to the cell layer, penetration across it, and diffusion away
from the layer to the balk solution of the acceptor tank in series.
Drug penetration across lipid layer includes passive diffusion, active
transport and diffusion through pores in the layer simulating paracellular
absorption.

The effective permeability coefficient, $P$, through a combination
of diffusional barriers and active transports is determined by the
following equation \citep{flynn1974mtp}:\begin{equation}
P^{-1}=P_{{\rm UWL}}^{-1}+(P_{{\rm pass}}+P_{{\rm act}}+P_{{\rm para}})^{-1},\label{eq:Papp}\end{equation}
where $P_{{\rm pass}}$, $P_{{\rm act}}$, $P_{{\rm para}}$are the
passive, active, paracellular permeabilities. $P_{UWL}$ is the effective
permeability of unstirred water layers (UWL) in the donor and acceptor
tanks:\[
P_{{\rm UWL}}^{-1}=P_{{\rm UWL},1}^{-1}+P_{{\rm UWL},2}^{-1}\]
The values of the permeabilities come from the Fick's law \begin{equation}
P_{{\rm UWL},i}=\frac{D_{{\rm UWL},i}}{h_{{\rm UWL},i}}\label{eq:Puwl}\end{equation}
where $D_{{\rm {\rm UWL},i}}$ and $h_{{\rm UWL},i}$ are the diffusion
coefficient and effective thickness of UWL on each side of the cell
monolayer. $D_{{\rm {\rm UWL},i}}$ can be approximated by the diffusional
coefficient in water, which varies within less than a single order
of magnitude for low molecular weight organic compounds \citep{wilke1955cdc,pade1997erc}.
For sufficiently dilute solutions $h_{UWL}$ is approximately constant.
Thus, $P_{UWL,i}$ and effective permeability of the UWLs, $P_{UWL}$
can be approximately treated constant for all low molecular weight
organic compounds.

Similarly to $P_{UWL,i}$, the drug diffusion through membrane, $P_{{\rm pass}}$,
can be estimated as:\begin{equation}
P_{{\rm pass}}=\frac{D_{M}}{h_{M}}D\label{eq:Pmem}\end{equation}
where $D_{M}$ is the diffusion coefficient in lipid, $h_{M}$ is
the thickness of the membrane, and $D$ is the octanol/water distribution
coefficient, i.e. the concentration ratio between aqueous and lipid
phases. And again, as a first approximation $D_{M}$ can be put to
a constant for various drug-like compounds, thus the proportionality
factor between $P_{pass}$ and $D$ can be considered as constant
for all low molecular weight organic compounds.

According to \citep{adson1994qad,adson1995pdw}, the paracellular
permeability, $P_{para}$, is a size-restricted diffusion within a
negative electrostatic force field. Normally it varies within a single
order of magnitude range \citep{adson1995pdw,pade1997erc} and hence
its variations can be neglected. In what follows we keep $P_{para}$
constant everywhere. The analysis of the experimental data at our
disposal proves that this is a very reasonable assumption indeed.

\begin{figure}
\includegraphics[width=0.9\columnwidth]{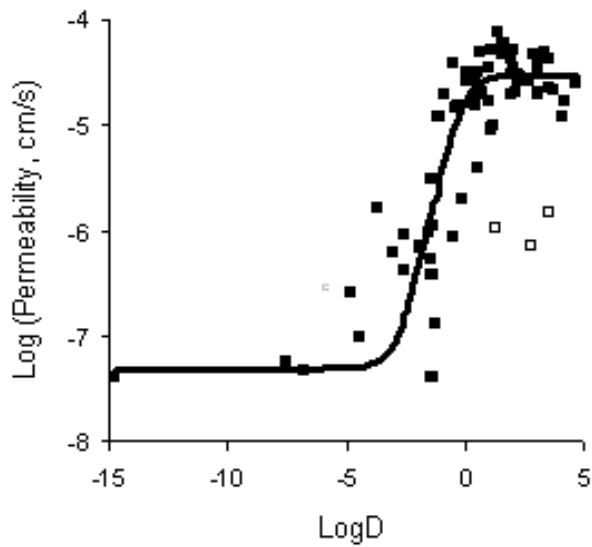}

\caption{Model of passive intestinal permeability. Intestinal permeability
of passively absorbed compounds (filled and hollow squares) is plotted
against LogD. Solid line is prediction of the model \ref{eq:Papp},
where $P_{{\rm act}}=0$ ${\rm cm\cdot s^{-1}}$, $P_{{\rm para}}={\rm const}$.
The parameter values see in the text.\label{fig:PassAbsorption}}

\end{figure}

To build up a model of active transport first we estimated parameters
of passive absorption ($P_{{\rm UWL}}$, $P_{{\rm para}}$, and $D_{M}/h_{M}$)
by fitting observed permeabilities for passively absorbed compounds
with Eq. \ref{eq:Papp} $P_{{\rm act}}=0$. Then these values were
frozen and observed permeability values for actively transported compounds
were fitted with Eq. \ref{eq:Papp} where $P_{act}$ was substituted
by the proposed model of active transport.

\subsection{Estimation the passive absorption model parameters.}

To estimate the parameters of the passive absorption we fitted observed
permeability values for drugs, which to the best of our knowledge
are passively absorbed, with Eq. \ref{eq:Papp} with no active transport
($P_{{\rm act}}=0$). Since the approximation contains only three
adjustable parameters of passive absorption, there was no need in
a large data set. Therefore we selected passively absorbed compounds
with Caco-2 permeabilities measured directly. This was done because
the observed $FA $ depend on experimental conditions and may include
effects of drug instability in the intestinal fluids, intestinal metabolism
and so on. On the contrary, the data on Caco-2 permeability are free
of those mentioned problems. On the other hand tight junctions of
Caco-2 cell monolayer are significantly less permeable \citep{matsson2005erd,balimane2006cip}
than in the intestine. 

Fig. \ref{fig:PassAbsorption} shows the logarithm of permeability
of passively absorbed drugs (both filled and hollow squares) plotted
against the logarithm of distribution coefficient. In accordance with
previously proposed model \citep{camenisch1998smp} the intestinal
permeability of passively absorbed drugs increases with increase in
the distribution coefficient and saturate at both low and high ends.
The increasing part reflects growth in membrane permeability with
increase in the distribution coefficient of a drug. The saturation
at upper limit reflects diffusional limitations imposed by UWLs for
highly lipophilic drugs. The saturation at low $\log D$ corresponds
to residual permeability through tight junctions. Solid line shows
fitting of experimental data with Eq. \ref{eq:Papp} where $P_{{\rm act}}=0$,
$P_{UWL}$, $P_{para}$, and $D_{M}/h_{M}$ are all assumed constant
for all the compounds. The best fit was achieved at the following
values of the model parameters: $P_{para}=5.01\times10^{-8}$ ${\rm cm\cdot s^{-1}}$,
$P_{UWL}=2.88\times10^{-5}$ ${\rm cm\cdot s^{-1}}$, $D_{M}/h_{M}=3.71\times10^{-5}$
${\rm cm\cdot s^{-1}}$. RMSD were $0.42$ log units.

The determined value of $P_{para}$ is slightly lower of experimental
estimations ranged from $10^{-7}\div10^{-6}$ ${\rm cm\cdot s^{-1}}$
\citep{adson1995pdw,pade1997erc}, while $P_{UWL}$ is in a good agreement
with the observed values at slow stirring rate ($5\times10^{-5}$
${\rm cm\cdot s^{-1}}$at 25 rpm , \citep{adson1995pdw}). Using the
commonly accepted value of the diffusion coefficient, $10^{-5}$ ${\rm cm}^{2}{\rm \times s}^{-1}$
\citep{wilke1955cdc,pade1997erc}, from Eq. \ref{eq:Puwl} we find
the effective thickness of UWLs :\[
h_{{\rm UWL}}\thicksim3\times10^{2}\,\mu m\]
that is in excellent agreement with previously estimated values between
$35$ and $800\,\mu m$, \citep{andreoli1971aul,fagerholm1995eee}.

\subsection{The active absorption model.}

Using literature data \citep{tsuji1996cmi,katsura2003iad,dobson2008cmc,vanbambeke2003aep,yee1997vpa,matsson2005erd,han2002tpd}
we selected 45 compounds from our database, which are reportedly absorbed
using active transport and to the best of our knowledge are not subject
to drug efflux \citep{yee1997vpa,vanbambeke2003aep,matsson2005erd,Fish2007ddi,katsura2003iad,VarmaM.V.S._mp0499196}.
To enrich the data set both the values of Caco-2 measured directly
and the calculated by $FA$ permeability values were used. If both
$FA$ and Caco-2 permeability were available for a given compound,
the value of $P$ calculated from the measured $FA$ was employed.
This is a reasonable approach, since Caco-2 cells are known to under
express some important drug transporters \citep{lennernas1996cba,chong1996vpt},
and thus Caco-2 permeability data for actively transported compounds
is less reliable than $FA$.

\begin{figure}
\includegraphics[width=0.9\columnwidth]{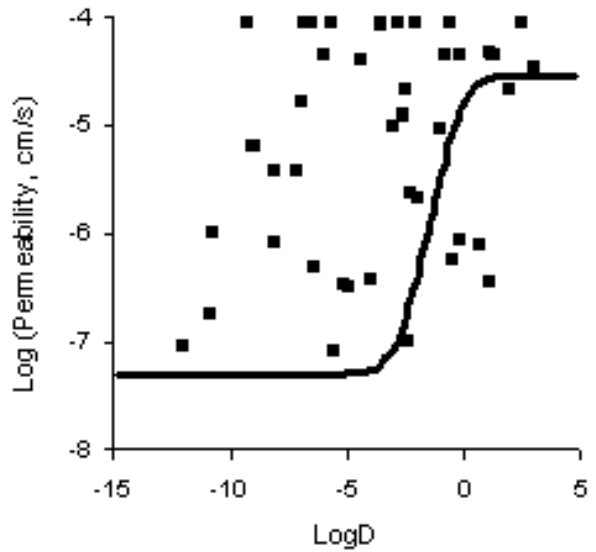}

\caption{Intestinal permeability of actively absorbed compounds. Pink points
are observed values. Solid line is the model of passive absorption.\label{fig:ActPerm}}

\end{figure}

Fig. \ref{fig:ActPerm} shows that permeability of the majority of
actively absorbed compounds were higher than predicted by the model
of passive absorption in accordance with existence of an additional
component of permeability. Nevertheless there were four outliers,
which permeability were substantially below passive permeability curve:
fosinopril, diphenhydramine, lobucavir, and cefuroxime axetil. We
will speculate about possible explanations of this in discussion.
For the rest of the compounds the total permeability exceeded passive
component from $0.06$ to $3.26$ log units and reached diffusion
limited rate. This means that the intensity of active transport varies
in the wide range and may be limited by drug diffusion to the membrane.

The carrier-mediated absorption (both active and passive) can be described
by Michaelis-Menten kinetics \citep{michaelis1913ki}:\begin{equation}
P_{act}=\Sigma\frac{n_{i}/\tau_{i}}{K_{D_{i}}+C}\label{eq:act_transp}\end{equation}
\begin{equation}
J_{act}=\Sigma n_{i}/\tau_{i}\frac{C}{K_{D_{i}}+C}\label{eq:act_flux}\end{equation}
where the summation occurs over all (the types of) transporters; $n_{i}$
is the amount of the $i$-th transporter molecules on the unit area
of membrane; $K_{D_{i}}$ is the dissociation constant of the i-th
transporter-ligand complex; $\tau_{i}$ is the time, required for
the transporter to bind and carry one molecule across the membrane;
$C$ - compound concentration. From Eq. \ref{eq:act_flux} it follows
that if $C\ll K_{D_{i}}$, then $J_{act}\rightarrow0$ (the compound
is passively absorbed). In the opposite case $C\gg K_{D_{i}}$ \[
J_{act}=\Sigma n_{i}/\tau_{i}\]
i.e. a compound actively absorbed and active component of a drug flux
is independent of drug concentration and determined only by the amount
of the protein-transporter and time, required for a transporter to
carry a ligand across membrane. Thus Eq. \ref{eq:act_flux} is similar
to a classifier, with threshold value C, which {}``selects'' between
the passive and the active transport options ({}``possibilities'').
Therefore it is natural to build a classifier model to identify proteins
that either participate in active transport directly, or have binding
site similar to that of a transporter.

To identify the proteins related to active absorption or with active
site force field similar to that of protein transporters we used all
drugs that are reported to be actively absorbed and have permeability
not less than predicted by passive model (41 compound). Besides, 71
passively absorbed drugs were used. %
\begin{figure}
\includegraphics[width=0.9\columnwidth]{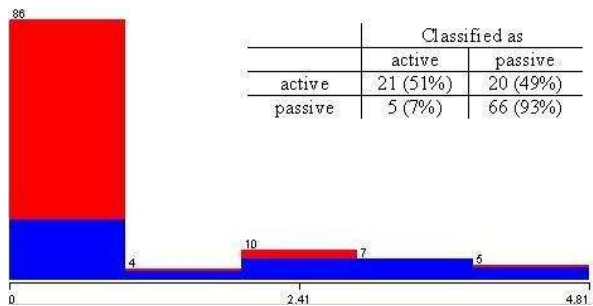}

\caption{Relation between affinity of a compound for human brain hexokinase
type I and intestinal absorption mechanism of the drug. Blue \textendash{}
actively absorbed compounds, red \textendash{} passively absorbed
compounds. X axis \textendash{} pKd value for the hexakinase. Y axis
\textendash{} the number of passively and actively absorbed compound.
The confusion matrix shows accuracy of prediction of active transport
with the help of human brain hexakinase.\label{fig:hexokinase}}

\end{figure}
We studied absorption-Kd relations for these drugs and proteins from
our set and identified a protein that correctly classified 78\% of
drugs between actively and passively transported. Fig. \ref{fig:hexokinase}
shows that practically all compounds with at least some small affinity
for the protein are actively absorbed. Using John Platt's sequential
minimal optimization algorithm for training a support vector classifier
implemented in Weka Data Mining Software we build up a classifier
model. The confusion matrix, as shown on Fig. \ref{fig:hexokinase},
shows that only 7\% ($5$ from $66$ passively transported compounds)
were mistakenly classified by the model as actively transported. These
outliers may in fact be false-positives, which affinity for the protein
was mistakenly calculated as high.

\begin{figure}

\includegraphics[width=0.9\columnwidth]{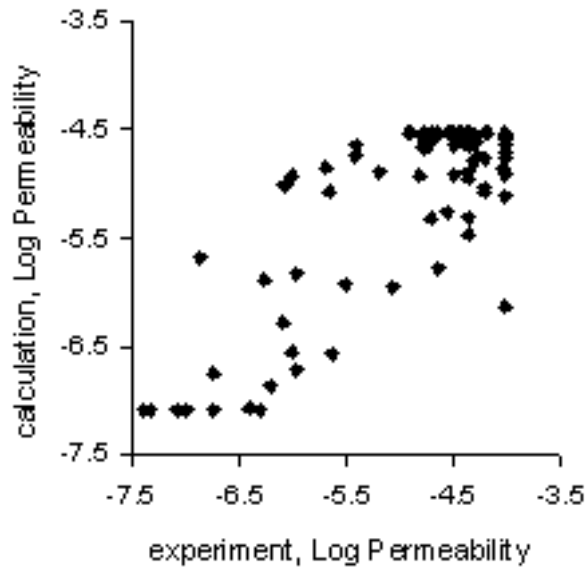}\caption{Permeability prediction for actively absorbed compounds. Predicted
permeability for correctly classified compounds is plotted against
experimental values. Both axis are in logarithmic scale.\label{fig:ActPermPrediction}}

\end{figure}

Fig. \ref{fig:ActPermPrediction} shows permeability prediction for
compounds that were correctly classified as active using Eqs. \ref{eq:Papp}
and \ref{eq:act_transp} where $n_{i}/\tau_{i}=3.2E-8\, M*cm^{-2}s^{-1}$
and $C=10E-30\, M*cm^{-3}$.%

Fig. \ref{fig:hexokinase} shows that there are a lot of actively
transported compounds among drugs with zero affinity for the protein.
This suggests that there should be other proteins that transport misclassified
drugs. The fact that we failed to find out them suggest that our protein
set misses some active site types that are important for intestinal
drug absorption. The confusion matrix shows that only 51\% (21 from
20 actively transported compounds) were classified as such. However,
if we consider only drugs with non-zero affinity for the protein (4
right bars on the histogram) 21 from 26 compounds (81\%) were actively
transported, showing that if a compound has affinity for the protein
it is most probably subject to active transport during intestinal
absorption. It is necessary to identify other proteins that classify
drugs as actively transported compounds.

In summary, we started from the premise that active drug transport
can be predicted by its affinity for some proteins, which in fact
not obligatory are transporters but have active site force field similar
to that of transporters. Using this approach and John Platt's sequential
minimal optimization algorithm for training a support vector classifier
we succeeded to identify a protein which affinity for drugs correlates
with the active absorption of these drugs in 81\% cases. This protein
can be used for estimation of the drug transport mechanism. The high
percent of actively absorbed compounds that the model mistakenly classifies
as passively absorbed indicates that there are other proteins that
transport these outliers. The fact that we failed to identify them
suggests that our current protein set misses some protein active site
types that are important for protein absorption and future steps should
be taken to enrich our protein set with such active sites. Nevertheless,
identification of the first protein that classifies compounds between
actively and passively absorbed shows that the proposed concept for
prediction of drug absorption is correct.

\section{Discussion}

At temps have been made to develop a theoretical model of oral absorption
and intestinal permeability \citep{camenisch1998smp,obata2005pod}.
Nevertheless, the models described only passive (trans and paracellular)
drug absorption, while active transport is an important part of it
\citep{dobson2008cmc}. In this paper we presented a novel approach
to in silico prediction of intestinal permeability for actively transported
compounds using binding data to some proteins. The developed model
(\ref{eq:Papp}) has the standard passive terms and includes an additional
active permeability term. When the last one is put to zero the model
reduces to a model of passive absorption. Fig. \ref{fig:PassAbsorption}
shows that the reduced (passive absorption) model fairly good describes
permeability of passively absorbed compounds. There were only three
apparent outliers (hollow squares): bupropion, bosentan, and remikiren.
If there were no errors in experimental data (either permeability
or distribution coefficient) we assumed that these compounds were
subjected to either efflux or intestinal metabolism. Fig \ref{fig:ActPerm}
shows that the majority of actively transported compounds were above
passive absorption curve, indicating existence of an additional component
of permeability, besides passive one. Nevertheless there were also
outliers, which permeability were substantially below passive permeability
curve: fosinopril, diphenhydramine, lobucavir, and cefuroxime axetil.
This can be due to inaccuracy in experimental data, or due to efflux
of the drug from the cell or due to intestinal metabolism, that does
not discovered yet.

The active absorption term in our model is determined by Michaelis-Menten
kinetics \citep{michaelis1913ki}. If a compound has high affinity
for important for active transport proteins it is predicted to be
actively absorbed otherwise the active component is low and the compound
is largely passively absorbed. The affinity data can be estimated
using molecular docking software. However, there are practically no
resolved structures of transporters, nevertheless there may be proteins
with similar force field in the active site that are solved and can
be used for docking. Indeed, we identified the first such protein
. It is human brain hexokinase. From 26 compounds with high calculated
affinity for the protein 21 (81\%) were actively transported. It is
not a transporter, in fact, however the correlation suggests that
that there is a transporter with similar active site force field.
Thus the transporter can be represented by this protein. Fig. \ref{fig:ActPermPrediction}shows
that the model using affinitiesfor this protein gives reasonably good
predictions for correctly classified compounds.

Further efforts should be made to identify the rest of the active
site types important for drug absorption in the intestine. Fig. \ref{fig:hexokinase}
shows that there are a lot of actively transported compounds among
drugs with zero affinity for the protein. This suggests that there
should be other active site types relevant to drug active transport.
The fact that we failed to identify them suggests that our protein
set misses some active site types that are important for intestinal
drug absorption. Thus the work should be continued in this direction.

\bibliographystyle{plain}
\bibliography{../Qrefs,Qrefs}

\begin{thebibliography}{10}

\bibitem{QUANTUM}
QUANTUM drug discovery software suit and services, http://www.q-pharm.com.

\bibitem{adson1995pdw}
A.~Adson, PS~Burton, TJ~Raub, CL~Barsuhn, KL~Audus, and NF~Ho.
\newblock {Passive diffusion of weak organic electrolytes across Caco-2 cell
  monolayers: uncoupling the contributions of hydrodynamic, transcellular, and
  paracellular barriers.}
\newblock {\em J Pharm Sci}, 84(10):1197--204, 1995.

\bibitem{adson1994qad}
A.~ADSON, TJ~RAUB, PS~BURTON, CL~BARSUHN, AR~HILGERS, KL~AUDUS, and NFH HO.
\newblock {Quantitative approaches to delineate paracellular diffusion in
  cultured epithelial cell monolayers}.
\newblock {\em Journal of pharmaceutical sciences}, 83(11):1529--1536, 1994.

\bibitem{agatonovickustrin2001tdm}
S.~Agatonovic-Kustrin, R.~Beresford, and A.P.M. Yusof.
\newblock {Theoretically-derived molecular descriptors important in human
  intestinal absorption}.
\newblock {\em Journal of Pharmaceutical and Biomedical Analysis},
  25(2):227--237, 2001.

\bibitem{andreoli1971aul}
T.E. Andreoli and S.L. Troutman.
\newblock {An Analysis of Unstirred Layers in Series with" Tight" and" Porous"
  Lipid Bilayer Membranes}.
\newblock {\em The Journal of General Physiology}, 57(4):464--478, 1971.

\bibitem{artursson1997ida}
P.~Artursson and RT~Borchardt.
\newblock {Intestinal drug absorption and metabolism in cell cultures: Caco-2
  and beyond.}
\newblock {\em Pharm Res}, 14(12):1655--8, 1997.

\bibitem{artursson1991cbo}
P.~Artursson and J.~Karlsson.
\newblock {Correlation between oral drug absorption in humans and apparent drug
  permeability coefficients in human intestinal epithelial (Caco-2) cells.}
\newblock {\em Biochem Biophys Res Commun}, 175(3):880--5, 1991.

\bibitem{artursson2001cme}
P.~Artursson, K.~Palm, and K.~Luthman.
\newblock {Caco-2 monolayers in experimental and theoretical predictions of
  drug transport}.
\newblock {\em Advanced Drug Delivery Reviews}, 46(1-3):27--43, 2001.

\bibitem{balimane2000cmu}
P.V. Balimane, S.~Chong, and R.A. Morrison.
\newblock {Current methodologies used for evaluation of intestinal permeability
  and absorption}.
\newblock {\em Journal of Pharmacological and Toxicological Methods},
  44(1):301--312, 2000.

\bibitem{balimane2006cip}
P.V. Balimane, Y.H. Han, and S.~Chong.
\newblock {Current industrial practices of assessing permeability and
  P-glycoprotein interaction}.
\newblock {\em The AAPS Journal}, 8(1):1--13, 2006.

\bibitem{balon1999dlp}
K.~Balon, BU~Riebesehl, and BW~Mueller.
\newblock {Drug Liposome Partitioning as a Tool for the Prediction of Human
  Passive Intestinal Absorption}.
\newblock {\em Pharmaceutical Research}, 16(6):882--888, 1999.

\bibitem{bergstrom2005spd}
C.A.S. Bergstrom.
\newblock {In silico Predictions of Drug Solubility and Permeability: Two
  Rate-limiting Barriers to Oral Drug Absorption}.
\newblock {\em Basic \& Clinical Pharmacology \& Toxicology}, 96(3):156--161,
  2005.

\bibitem{camenisch1998epp}
G.~Camenisch, J.~Alsenz, H.~van~de Waterbeemd, and G.~Folkers.
\newblock {Estimation of permeability by passive diffusion through Caco-2 cell
  monolayers using the drugs' lipophilicity and molecular weight}.
\newblock {\em European Journal of Pharmaceutical Sciences}, 6(4):313--319,
  1998.

\bibitem{camenisch1998smp}
G.~Camenisch, G.~Folkers, and H.~van~de Waterbeemd.
\newblock {Shapes of membrane permeability--lipophilicity curves: Extension of
  theoretical models with an aqueous pore pathway}.
\newblock {\em European Journal of Pharmaceutical Sciences}, 6(4):321--329,
  1998.

\bibitem{castillogarit2007eap}
JA~Castillo-Garit, Y.~Marrero-Ponce, F.~Torrens, and R.~Garcia-Domenech.
\newblock {Estimation of ADME properties in drug discovery: Predicting Caco-2
  cell permeability using atom-based stochastic and non-stochastic linear
  indices.}
\newblock {\em J Pharm Sci}, 97(5):1946--76, 2008.

\bibitem{chong1996vpt}
S.~Chong, SA~Dango, KM~Soucek, and RA~Morrison.
\newblock {In vitro permeability through caco-2 cells is not quantitatively
  predictive of in vivo absorption for peptide-like drugs absorbed via the
  dipeptide transporter system}.
\newblock {\em Pharmaceutical research}, 13(1):120--123, 1996.

\bibitem{clark1999rcp}
D.E. Clark.
\newblock {Rapid calculation of polar molecular surface area and its
  application to the prediction of transport phenomena. 1. Prediction of
  intestinal absorption}.
\newblock {\em Journal of Pharmaceutical Sciences}, 88(8):807--814, 1999.

\bibitem{dobson2008cmc}
P.D. Dobson and D.B. Kell.
\newblock {Carrier-mediated cellular uptake of pharmaceutical drugs: an
  exception or the rule?}
\newblock {\em Nature Reviews Drug Discovery}, 7(3):205, 2008.

\bibitem{eddershaw2000app}
P.J. Eddershaw, A.P. Beresford, and M.K. Bayliss.
\newblock {ADME/PK as part of a rational approach to drug discovery}.
\newblock {\em Drug Discovery Today}, 5(9):409--414, 2000.

\bibitem{ertl2000fcm}
P.~Ertl, B.~Rohde, and P.~Selzer.
\newblock {Fast Calculation of Molecular Polar Surface Area as a Sum of
  Fragment-Based Contributions and Its Application to the Prediction of Drug
  Transport Properties}.
\newblock {\em JOURNAL OF MEDICINAL CHEMISTRY}, 43(20):3714--3717, 2000.

\bibitem{fagerholm1995eee}
U.~Fagerholm and H.~Lennern{\"a}s.
\newblock {Experimental estimation of the effective unstirred water layer
  thickness in the human jejunum, and its importance in oral drug absorption}.
\newblock {\em European Journal of Pharmaceutical Sciences}, 3(5):247--253,
  1995.

\bibitem{fedichev-2006}
PO~Fedichev and LI~Men'shikov.
\newblock {Long-Range Order and Interactions of Macroscopic Objects in Polar
  Liquids}.
\newblock {\em Arxiv preprint cond-mat/0601129}, 2006.

\bibitem{fedichev2008fep}
PO~Fedichev and LI~Menshikov.
\newblock {Ferro-electric phase transition in a polar liquid and the nature
  of$\backslash$ lambda-transition in supercooled water}.
\newblock {\em eprint arXiv: 0808.0991}, 2008.

\bibitem{Fish2007ddi}
J.~Fish.
\newblock {Drug-Drug Interactions. A guide to identifying and managing
  important drug interactions}.
\newblock {\em Journal of the Pharmacy Society of Wisconsin},
  July/August:16--25, 2007.

\bibitem{fliri2005bsa}
A.F. Fliri, W.T. Loging, P.F. Thadeio, and R.A. Volkmann.
\newblock {Biological spectra analysis: Linking biological activity profiles to
  molecular structure}.
\newblock {\em Proceedings of the National Academy of Sciences},
  102(2):261--266, 2005.

\bibitem{fliri2005bam}
A.F. Fliri, W.T. Loging, P.F. Thadeio, and R.A. Volkmann.
\newblock {Biospectra analysis: model proteome characterizations for linking
  molecular structure and biological response}.
\newblock {\em J. Med. Chem}, 48(22):6918--6925, 2005.

\bibitem{flynn1974mtp}
GL~Flynn, SH~Yalkowsky, and TJ~Roseman.
\newblock {Mass transport phenomena and models: theoretical concepts.}
\newblock {\em J Pharm Sci}, 63(4):479--510, 1974.

\bibitem{ghuloum1999mhn}
A.M. Ghuloum, C.R. Sage, and A.N. Jain.
\newblock {Molecular Hashkeys: A Novel Method for Molecular Characterization
  and Its Application for Predicting Important Pharmaceutical Properties of
  Molecules}.
\newblock {\em JOURNAL OF MEDICINAL CHEMISTRY}, 42:1739--1748, 1999.

\bibitem{guangli2006pcp}
M.~Guangli and C.~Yiyu.
\newblock {Predicting Caco-2 Permeability Using Support Vector Machine and
  Chemistry Development Kit}.
\newblock {\em J Pharm Pharm Sci}, 9(2):210--21, 2006.

\bibitem{gunturi2007sam}
SB~Gunturi and R.~Narayanan.
\newblock {In Silico ADME Modeling 3: Computational Models to Predict Human
  Intestinal Absorption Using Sphere Exclusion and kNN QSAR Methods}.
\newblock {\em QSAR AND COMBINATORIAL SCIENCE}, 26(5):653--668, 2007.

\bibitem{han2002tpd}
H.K. Han and G.L. Amidon.
\newblock {Targeted prodrug design to optimize drug delivery}.
\newblock {\em The AAPS Journal}, 2(1):48--58, 2002.

\bibitem{hou2007aed2}
T.~Hou, J.~Wang, and Y.~Li.
\newblock {ADME Evaluation in Drug Discovery. 8. The Prediction of Human
  Intestinal Absorption by a Support Vector Machine}.
\newblock {\em J Chem Inf Model}, 47(6):2408--15, 2007.

\bibitem{hou2006rac}
T.~Hou, J.~Wang, W.~Zhang, W.~Wang, and X.~Xu.
\newblock {Recent Advances in Computational Prediction of Drug Absorption and
  Permeability in Drug Discovery}.
\newblock {\em Current Medicinal Chemistry}, 13(22):2653--2667, 2006.

\bibitem{hou2007aed1}
T.~Hou, J.~Wang, W.~Zhang, and X.~Xu.
\newblock {ADME Evaluation in Drug Discovery. 7. Prediction of Oral Absorption
  by Correlation and Classification}.
\newblock {\em J. Chem. Inf. Model}, 47(1):208--218, 2007.

\bibitem{hunter1997isd}
J.~Hunter and B.H. Hirst.
\newblock {Intestinal secretion of drugs. The role of P-glycoprotein and
  related drug efflux systems in limiting oral drug absorption}.
\newblock {\em Advanced Drug Delivery Reviews}, 25(2-3):129--157, 1997.

\bibitem{ingels2003bpa}
F.M. Ingels and P.F. Augustijns.
\newblock {Biological, pharmaceutical, and analytical considerations with
  respect to the transport media used in the absorption screening system,
  Caco-2}.
\newblock {\em Journal of pharmaceutical sciences}, 92(8):1545--1558, 2003.

\bibitem{irvine1999mmd}
JD~Irvine, L.~Takahashi, K.~Lockhart, J.~Cheong, JW~Tolan, HE~Selick, and
  JR~Grove.
\newblock {MDCK(Madin-Darby canine kidney) cells: A tool for membrane
  permeability screening}.
\newblock {\em Journal of pharmaceutical sciences}, 88(1):28--33, 1999.

\bibitem{kansy1998pht}
M.~KANSY, F.~SENNER, and K.~GUBERNATOR.
\newblock {Physicochemical high throughput screening: Parallel artificial
  membrane permeation assay in the description of passive absorption
  processes}.
\newblock {\em Journal of medicinal chemistry(Print)}, 41(7):1007--1010, 1998.

\bibitem{katsura2003iad}
T.~KATSURA and K.~INUI.
\newblock {Intestinal Absorption of Drugs Mediated by Drug Transporters:
  Mechanisms and Regulation}.
\newblock {\em Drug Metabolism and Pharmacokinetics}, 18(1):1--15, 2003.

\bibitem{klopman2002aec}
G.~Klopman, L.R. Stefan, and R.D. Saiakhov.
\newblock {ADME evaluation 2. A computer model for the prediction of intestinal
  absorption in humans}.
\newblock {\em European Journal of Pharmaceutical Sciences}, 17(4-5):253--263,
  2002.

\bibitem{lennernas1996cba}
H.~Lennern{\"a}s, K.~Palm, U.~Fagerholm, and P.~Artursson.
\newblock {Comparison between active and passive drug transport in human
  intestinal epithelial (caco-2) cells in vitro and human jejunum in vivo}.
\newblock {\em International Journal of Pharmaceutics}, 127(1):103--107, 1996.

\bibitem{liang2000mta}
E.~Liang, J.~Proudfoot, and M.~Yazdanian.
\newblock {Mechanisms of Transport and Structure-Permeability Relationship of
  Sulfasalazine and Its Analogs in Caco-2 Cell Monolayers}.
\newblock {\em Pharmaceutical Research}, 17(10):1168--1174, 2000.

\bibitem{linnankoski2006cpo}
J.~Linnankoski, J.~Maekelae, V.P. Ranta, A.~Urtti, and M.~Yliperttula.
\newblock {Computational prediction of oral drug absorption based on absorption
  rate constants in humans}.
\newblock {\em J. Med. Chem}, 49(12):3674--3681, 2006.

\bibitem{loging2007hte}
W.~Loging, L.~Harland, B.~Williams-Jones, et~al.
\newblock {High-throughput electronic biology: mining information for drug
  discovery}.
\newblock {\em Nature Reviews Drug Discovery}, 6:220--230, 2007.

\bibitem{martinez2004cnn}
A.~Martinez.
\newblock {CODES/Neural Network Model: a Useful Tool for in Silico Prediction
  of Oral Absorption and Blood-Brain Barrier Permeability of Structurally
  Diverse Drugs}.
\newblock {\em QSAR Comb. Sci}, page~23, 2004.

\bibitem{matsson2005erd}
P.~Matsson, C.A.S. Bergstr{\"o}m, N.~Nagahara, S.~Tavelin, U.~Norinder, and
  P.~Artursson.
\newblock {Exploring the role of different drug transport routes in
  permeability screening}.
\newblock {\em J. Med. Chem}, 48(2):604--613, 2005.

\bibitem{matsson2007gdi}
P.~Matsson, G.~Englund, G.~Ahlin, C.A.S. Bergstr{\"o}m, U.~Norinder, and
  P.~Artursson.
\newblock {A Global Drug Inhibition Pattern for the Human ATP-Binding Cassette
  Transporter Breast Cancer Resistance Protein (ABCG2)[boxs]}.
\newblock {\em Journal of Pharmacology and Experimental Therapeutics},
  323(1):19--30, 2007.

\bibitem{michaelis1913ki}
L.~Michaelis and M.L. Menten.
\newblock {Die Kinetik der Invertinwirkung}.
\newblock {\em Biochem. Z}, 49(333):148, 1913.

\bibitem{nordqvist2004gmp}
A.~NORDQVIST, J.~NILSSON, T.~LINDMARK, A.~ERIKSSON, P.~GARBERG, and M.~KIHLEN.
\newblock {A general model for prediction of Caco-2 cell permeability}.
\newblock {\em QSAR and combinatorial science(Print)}, 23(5):303--310, 2004.

\bibitem{obata2005pod}
K.~Obata, K.~Sugano, R.~Saitoh, A.~Higashida, Y.~Nabuchi, M.~Machida, and
  Y.~Aso.
\newblock {Prediction of oral drug absorption in humans by theoretical passive
  absorption model}.
\newblock {\em International Journal of Pharmaceutics}, 293(1-2):183--192,
  2005.

\bibitem{oprea1999tmm}
T.I. Oprea and J.~Gottfries.
\newblock {Toward minimalistic modeling of oral drug absorption}.
\newblock {\em Journal of Molecular Graphics and Modelling}, 17(5-6):261--274,
  1999.

\bibitem{pade1997erc}
V.~Pade and S.~Stavchansky.
\newblock {Estimation of the Relative Contribution of the Transcellular and
  Paracellular Pathway to the Transport of Passively Absorbed Drugs in the
  Caco-2 Cell Culture Model}.
\newblock {\em Pharmaceutical Research}, 14(9):1210--1215, 1997.

\bibitem{palm1997pms}
K.~Palm, P.~Stenberg, K.~Luthman, and P.~Artursson.
\newblock {Polar Molecular Surface Properties Predict the Intestinal Absorption
  of Drugs in Humans}.
\newblock {\em Pharmaceutical Research}, 14(5):568--571, 1997.

\bibitem{parrott2002pia}
N.~Parrott and T.~Lave.
\newblock {Prediction of intestinal absorption: comparative assessment of
  gastroplus and idea}.
\newblock {\em European Journal of Pharmaceutical Sciences}, 17(1-2):51--61,
  2002.

\bibitem{ponce2004ntd}
Y.M. Ponce, M.A.C. Perez, V.R. Zaldivar, H.G. Diaz, and F.~Torrens.
\newblock {A new topological descriptors based model for predicting intestinal
  epithelial transport of drugs in caco-2 cell culture.}
\newblock {\em J Pharm Pharm Sci}, 7:186--99, 2004.

\bibitem{raevsky2000qed}
O.A. Raevsky, V.I. Fetisov, E.P. Trepalina, J.W. McFarland, and K.J. Schaper.
\newblock {Quantitative Estimation of Drug Absorption in Humans for Passively
  Transported Compounds on the Basis of Their Physico-chemical Parameters}.
\newblock {\em Quantitative Structure-Activity Relationships}, 19(4):366--374,
  2000.

\bibitem{raevsky2002nap}
O.A. Raevsky, K.J. Schaper, P.~Artursson, and J.W. McFarland.
\newblock {A Novel Approach for Prediction of Intestinal Absorption of Drugs in
  Humans based on Hydrogen Bond Descriptors and Structural Similarity}.
\newblock {\em QUANTITATIVE STRUCTURE ACTIVITY RELATIONSHIPS},
  20(5/6):402--413, 2002.

\bibitem{sanghvi2003ppi}
T.~Sanghvi, N.~Ni, M.~Mayersohn, and S.H. Yalkowsky.
\newblock {Predicting Passive Intestinal Absorption Using A Single Parameter}.
\newblock {\em QSAR and Combinatorial Science}, 22(2):247--257, 2003.

\bibitem{schwede2003sma}
T.~Schwede et~al.
\newblock {SWISS-MODEL: an automated protein homology-modeling server}.
\newblock {\em Nucleic Acids Research}, 31(13):3381--3385, 2003.

\bibitem{stenberg2001eac}
P.~Stenberg, U.~Norinder, K.~Luthman, and P.~Artursson.
\newblock {Experimental and Computational Screening Models for the Prediction
  of Intestinal Drug Absorption}.
\newblock {\em JOURNAL OF MEDICINAL CHEMISTRY}, 44(12):1927--1937, 2001.

\bibitem{subramanian2006cam}
G.~Subramanian and D.B. Kitchen.
\newblock {Computational approaches for modeling human intestinal absorption
  and permeability}.
\newblock {\em Journal of Molecular Modeling}, 12(5):577--589, 2006.

\bibitem{tavelin1999cie}
S.~Tavelin, V.~Milovic, G.~Ocklind, S.~Olsson, and P.~Artursson.
\newblock {A Conditionally Immortalized Epithelial Cell Line for Studies of
  Intestinal Drug Transport}.
\newblock {\em Journal of Pharmacology and Experimental Therapeutics},
  290(3):1212--1221, 1999.

\bibitem{tsuji1996cmi}
A.~Tsuji and I.~Tamai.
\newblock {Carrier-Mediated Intestinal Transport of Drugs}.
\newblock {\em Pharmaceutical Research}, 13(7):963--977, 1996.

\bibitem{vanbambeke2003aep}
F.~Van~Bambeke, J.M. Michot, and PM~Tulkens.
\newblock {Antibiotic efflux pumps in eukaryotic cells: occurrence and impact
  on antibiotic cellular pharmacokinetics, pharmacodynamics and
  toxicodynamics.}
\newblock {\em Journal of Antimicrobial Chemotherapy}, 51(5):1067, 2003.

\bibitem{vandewaterbeemd2001pbd}
H.~van De~Waterbeemd, DA~Smith, K.~Beaumont, and DK~Walker.
\newblock {Property-based design: optimization of drug absorption and
  pharmacokinetics.}
\newblock {\em J Med Chem}, 44(9):1313--33, 2001.

\bibitem{VarmaM.V.S._mp0499196}
M.V.S. Varma, K.~Sateesh, and R.~Panchagnula.
\newblock Functional role of p-glycoprotein in limiting intestinal absorption
  of drugs: Contribution of passive permeability to p-glycoprotein mediated
  efflux transport.
\newblock {\em Molecular Pharmaceutics}, 2(1):12--21, 2005.

\bibitem{wessel1998phi}
M.D. Wessel, P.C. Jurs, J.W. Tolan, and S.M. Muskal.
\newblock {Prediction of Human Intestinal Absorption of Drug Compounds from
  Molecular Structure}.
\newblock {\em JOURNAL OF CHEMICAL INFORMATION AND COMPUTER SCIENCES},
  38:726--735, 1998.

\bibitem{PDB:Westbrook2003}
J.~Westbrook, Z.~Feng, L.~Chen, H.~Yang, and H.~M. Berman.
\newblock The protein data bank and structural genomics.
\newblock {\em Nucleic Acids Res}, 31(1):489--491, January 2003.

\bibitem{wilke1955cdc}
CR~Wilke and P.~Chang.
\newblock {Correlation of diffusion coefficients in dilute solutions}.
\newblock {\em AIChE Journal}, 1(2):264--270, 1955.

\bibitem{WEKA:witten2005dmp}
I.H. Witten and E.~Frank.
\newblock {\em {Data Mining: Practical Machine Learning Tools and Techniques}}.
\newblock Morgan Kaufmann, 2005.

\bibitem{yamazaki2004cpp}
K.~Yamazaki and M.~Kanaoka.
\newblock {Computational prediction of the plasma protein-binding percent of
  diverse pharmaceutical compounds}.
\newblock {\em JOURNAL OF PHARMACEUTICAL SCIENCES}, 93(6):1480--1494, 2004.

\bibitem{yazdanian1998cpa}
M.~Yazdanian, SL~Glynn, JL~Wright, and A.~Hawi.
\newblock {Correlating Partitioning and Caco-2 Cell Permeability of
  Structurally Diverse Small Molecular Weight Compounds}.
\newblock {\em Pharmaceutical Research}, 15(9):1490--1494, 1998.

\bibitem{yee1997vpa}
S.~Yee.
\newblock {In Vitro Permeability Across Caco-2 Cells (Colonic) Can Predict In
  Vivo (Small Intestinal) Absorption in Man: Fact or Myth}.
\newblock {\em Pharmaceutical Research}, 14(6):763--766, 1997.

\bibitem{yu1999caa}
L.X. Yu and G.L. Amidon.
\newblock {A compartmental absorption and transit model for estimating oral
  drug absorption}.
\newblock {\em International Journal of Pharmaceutics}, 186(2):119--125, 1999.

\bibitem{zhao2002rls}
Y.H. Zhao, M.H. Abraham, J.~Le, A.~Hersey, C.N. Luscombe, G.~Beck,
  B.~Sherborne, and I.~Cooper.
\newblock {Rate-Limited Steps of Human Oral Absorption and QSAR Studies}.
\newblock {\em Pharmaceutical Research}, 19(10):1446--1457, 2002.

\end{thebibliography}

\end{document}